\documentstyle[aps,prb]{revtex}

\begin{document}

\input epsf.sty
\twocolumn[\hsize\textwidth\columnwidth\hsize\csname %
@twocolumnfalse\endcsname

\draft

\widetext

\title{Direct Observation of a One Dimensional Static Spin Modulation in Insulating 
La$_{1.95}$Sr$_{0.05}$CuO$_{4}$}

\author{S. Wakimoto\footnote{Also at Department of Physics, Brookhaven 
National Laboratory, Upton, NY 11973}, R. J. Birgeneau, M. A. Kastner and Y. S. Lee}
\address{Department of Physics and Center for Materials Science and Engineering, 
Massachusetts Institute of Technology, 
Cambridge, Massachusetts 02139}
\author{R. Erwin, P. M. Gehring and S. H. Lee\footnote{Also at University 
of Maryland, College Park, MD 20742}}
\address{National Institute of Standards and Technology, NCNR, Gaithersburg, Maryland 20889}
\author{M. Fujita and K. Yamada}
\address{Institute for Chemical Research, Kyoto University, Gokasho, 
Uji 610-0011, Japan}
\author{Y. Endoh and K. Hirota}
\address{Department of Physics, Tohoku University, Sendai 980-8578, Japan}
\author{G. Shirane}
\address{Brookhaven National Laboratory, Upton, New York 11973}

\date{\today}
\maketitle


\begin{abstract}

We report the results of an extensive elastic neutron scattering study of the 
incommensurate (IC) {\it static} spin correlations in La$_{1.95}$Sr$_{0.05}$CuO$_{4}$
which is an insulating spin glass at low temperatures. 
Recent work by Wakimoto {\it et al.} has revealed the
presence of new two dimensional satellite peaks in La$_{1.95}$Sr$_{0.05}$CuO$_{4}$ at positions
rotated by $\sim 45^{\circ}$ in reciprocal space from those found in superconducting samples.
The present neutron scattering experiments on the same $x=0.05$ crystal employ a
narrower instrumental $Q$-resolution and thereby have revealed that the crystal has only two, 
rather than four, orthorhombic
twins at low temperatures with relative populations of 2:1.  
This has made possible 
the precise characterization of the IC elastic peaks around $(1,\ 0,\ 0)$ and
$(0,\ 1,\ 0)$ (orthorhombic notation) in each domain separately.  
We find that, in a single twin,
only two satellites are observed at $(1,\ \pm 0.064,\ L)_{ortho}$ and 
$(0,\ 1\pm0.064,\ L)_{ortho}$, that is, the modulation vector is only along the orthorhombic $b^{*}$-axis.
This demonstrates unambiguously that La$_{1.95}$Sr$_{0.05}$CuO$_{4}$ has a 
one-dimensional rather than two-dimensional static diagonal spin modulation 
at low temperatures, consistent with certain stripe models.
From the L-dependence we conclude that the spin correlations are predominantly two dimensional.
We have also reexamined the
$x=0.04$ crystal that previously was 
reported to show a single commensurate peak. By mounting the sample in the
$(H,\ K,\ 0)$ zone, we have discovered that the $x=0.04$ sample in fact has the 
same IC structure as  
the $x=0.05$ sample.
The incommensurability parameter $\delta$ for $x=0.04$ and $0.05$, 
where $\delta$ is the distance from (1/2, 1/2) in tetragonal reciprocal lattice units,
follows the linear relation $\delta \simeq x$. 
These results demonstrate that the insulator to superconductor transition in the under doped 
regime $(0.05 \leq x \leq 0.06)$ in La$_{2-x}$Sr$_{x}$CuO$_{4}$ is coincident with a 
transition from diagonal to collinear static stripes at low temperatures thereby
evincing the intimate coupling between the one dimensional 
spin density modulation and the superconductivity.

\end{abstract}


\phantom{.}
]
\narrowtext

%
%

\section{Introduction}
\label{sec_intro}

\begin{figure}
\centerline{\epsfxsize=3in\epsfbox{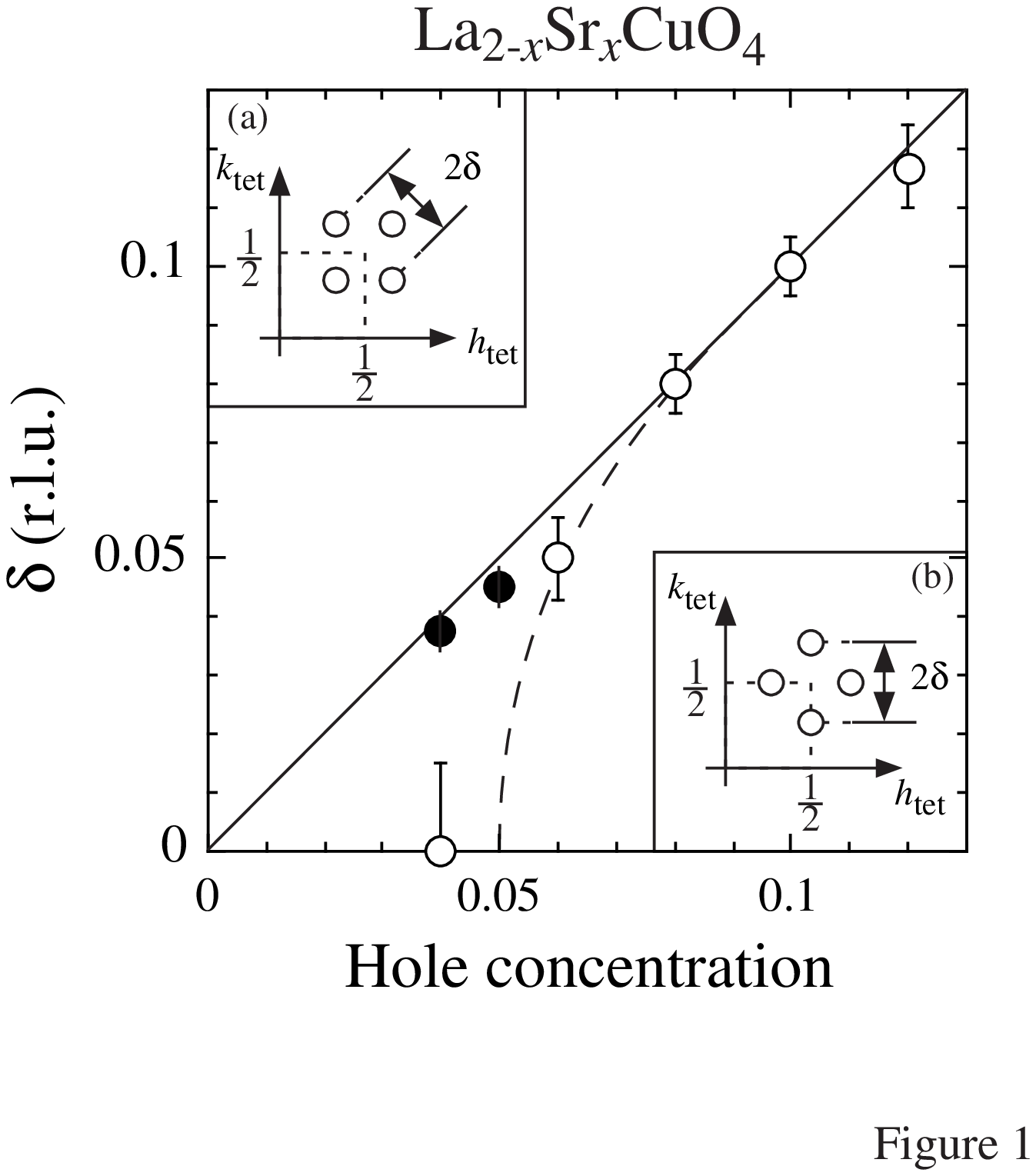}}
\caption{Hole concentration dependence of the incommensurate amplitude $\delta$.
$\delta$ is defined as the distance between the IC peaks and the $(\pi,\ \pi)$ position
using the in-plane reciprocal lattice unit of the tetragonal $I4/mmm$ structure.
Open circles indicate the data for the inelastic IC peaks reported by 
Yamada {\it et al.}~\cite{K.Yamada_98}
Closed circles are data from the present study.
The solid line corresponds to $\delta=x$.}
\label{fig_incomme}
\end{figure}

In current studies of the microscopic physics of high-$T_{C}$ superconductivity, 
the relationship between the magnetism and the superconductivity has become 
one of the central foci of attention.~\cite{M.A.Kastner_98} 
Most especially, La$_{2}$CuO$_{4}$ 
and related compounds have been studied in detail since they are
among the simplest of the high-T$_{C}$ materials with 
single CuO$_{2}$ planes composed of square Cu$^{2+}$ lattices. 
In the superconducting hole concentration range 
$x \stackrel{>}{\sim} 0.06$, the La$_{2-x}$Sr$_{x}$CuO$_{4}$ (LSCO) system exhibits
two dimensional dynamic 
magnetic correlations which give rise to incommensurate (IC) peaks at 
$(\frac{1}{2}\pm\delta, \frac{1}{2})$ and $(\frac{1}{2}, \frac{1}{2}\pm\delta)$ 
in the tetragonal square lattice notation
shown in the inset (b) in 
Fig.~\ref{fig_incomme}.~\cite{M.A.Kastner_98,Yoshizawa_88,Bob_89,S.W.Cheong_91,T.Mason_93,M.Matsuda_94,K.Yamada_98}

Spin fluctuations in the superconducting concentration range were first reported 
in Ref.2.
The incommensurability of the spin fluctuations was discovered independently by
Yoshizawa {\it et al.}~\cite{Yoshizawa_88} and Birgeneau {\it et al.}~\cite{Bob_89} 
and the explicit geometry of the fluctuations
was subsequently elucidated by Cheong {\it et al.}~\cite{S.W.Cheong_91}
The IC peak positions correspond to spin modulation vectors that are parallel 
to the two tetragonal axes. 
We refer to this as the {\it collinear} spin density wave orientation.
In seminal work, Yamada {\it et al.}~\cite{K.Yamada_98}
discovered a remarkably simple relationship between the incommensurability $\delta$ 
and the doping concentration $x$; $\delta$ obeys a linear relation, 
$\delta \simeq x$, for $0.06 \leq x \leq 0.12$, as indicated in Fig.~\ref{fig_incomme}
while for larger $x$, $\delta$ saturates near 1/8.
Very recently, Wakimoto {\it et al.}~\cite{waki_rapid} 
discovered a new class of satellite peaks in the insulating spin glass
compound, La$_{1.95}$Sr$_{0.05}$CuO$_{4}$, at the positions schematically illustrated in the inset (a)
of Fig.~\ref{fig_incomme}.
The same geometry for the elastic IC peaks has been reported in the insulating 
La$_{2-x}$Sr$_{x}$NiO$_{4}$ compound.~\cite{Hayden_92,T.traNi_96,notedelta}
In both compounds, the spin modulation vectors are 
parallel to the orthorhombic axes, which in turn are rotated by 45$^{\circ}$ from those of 
the superconducting LSCO samples. 
These facts suggest a strong correlation 
between the transport properties and the direction of the magnetic modulation. 

However, to-date in high-$T_{C}$ materials there is only limited information 
on the correlation between
the detailed crystal structure and the magnetic IC structure.~\cite{M.A.Kastner_98}
In the low-temperature orthorhombic (LTO) phase, the static IC structures 
found in superconducting LSCO~\cite{T.Suzuki_98,Kimura_99} 
and La$_{2}$CuO$_{4.12}$~\cite{Lee_99} have spin
modulations which are approximately along the
$a_{tet}$ and $b_{tet}$-axes which are almost equivalent, since both modulation vectors 
are at about 45$^{\circ}$ with respect to the CuO$_{6}$ tilt direction.
The spin correlations typically extend
in the $c$-direction to between two and three CuO$_{2}$ planes.
Importantly, for the $x=0.05$ material, IC states with the modulation vectors along the $a_{ortho}$ and 
$b_{ortho}$-axes are quite distinct, since one is perpendicular and the other is
parallel to the CuO$_{6}$ tilt axis.
Therefore, it is essential that the IC peaks be investigated
taking into account the explicit orthorhombic structure.

With the above as motivation, we have carried out an extensive study of the static IC peaks 
for $x=0.05$ using the same sample as reported on previously.~\cite{waki_rapid}
In this study we have taken into account particularly the fact that the $x=0.05$ sample has a 
twin structure due to the orthorhombic distortion.
Fortunately, our $x=0.05$ sample 
contains only two rather than all four possible twins. (See section \ref{sec_geometry}.) 
Furthermore, in each twin the observed spin modulation peaks are
in a direction parallel to the $b^{*}_{ortho}$-axis.
Importantly, this means that the spin modulation is one dimensional rather than two 
dimensional, consistent with certain stripe models. (See section \ref{sec_result}.)

We also have reexamined a crystal of  
La$_{1.96}$Sr$_{0.04}$CuO$_{4}$ in which we previously observed only a broad 
commensurate peak.~\cite{waki_rapid}
In light of the results for $x=0.05$, we have made measurements using the optimal 
orientation of the crystal and find
the same type of IC peaks as those in the $x=0.05$ sample.

Recently,
Suzuki {\it et al.}~\cite{T.Suzuki_98} and 
Kimura {\it et al.}~\cite{Kimura_99} have observed elastic IC peaks
oriented approximately along the tetragonal axes $\hat{a}_{tet}$ and $\hat{b}_{tet}$ in 
La$_{1.88}$Sr$_{0.12}$CuO$_{4}$ in the LTO phase
with an onset temperature of $\sim T_{C}$ (superconducting). 
Subsequently, 
IC elastic peaks have been observed in electrochemically-oxidized stage~-~4
La$_{2}$CuO$_{4+\delta}$ by Lee {\it et al.}~\cite{Lee_99} 
with the onset also at $T_{C}$ and 
\linebreak
\noindent
they established that
the orientation of the IC peaks deviates subtly from 
collinearity with the tetragonal crystal axes. 
The same deviation has also recently been confirmed in La$_{1.88}$Sr$_{0.12}$CuO$_{4}$.~\cite{Kimura_99} 
Our new results together with these previous results naturally lead to
a more coherent description of the IC peaks of the LSCO system 
over a wide hole concentration range using polar coordinates. (See section \ref{sec_polar})
The relation of the IC spin modulation to the transport properties
as well as the crystal structure will be discussed in the context of the stripe model.
Most notably, previous work has revealed that the insulator-superconductor transition 
around $x \simeq 0.05$ corresponds to a commensurate-incommensurate transition in the 
{\it instantaneous} spin correlations~\cite{M.A.Kastner_98,Bob_89} 
whereas our results reveal that the onset of superconductivity coincides with 
a diagonal-collinear
transition in the modulation vector of the {\it static} spin correlations.

\section{Scattering geometry}
\label{sec_geometry}

\begin{figure}
\centerline{\epsfxsize=3in\epsfbox{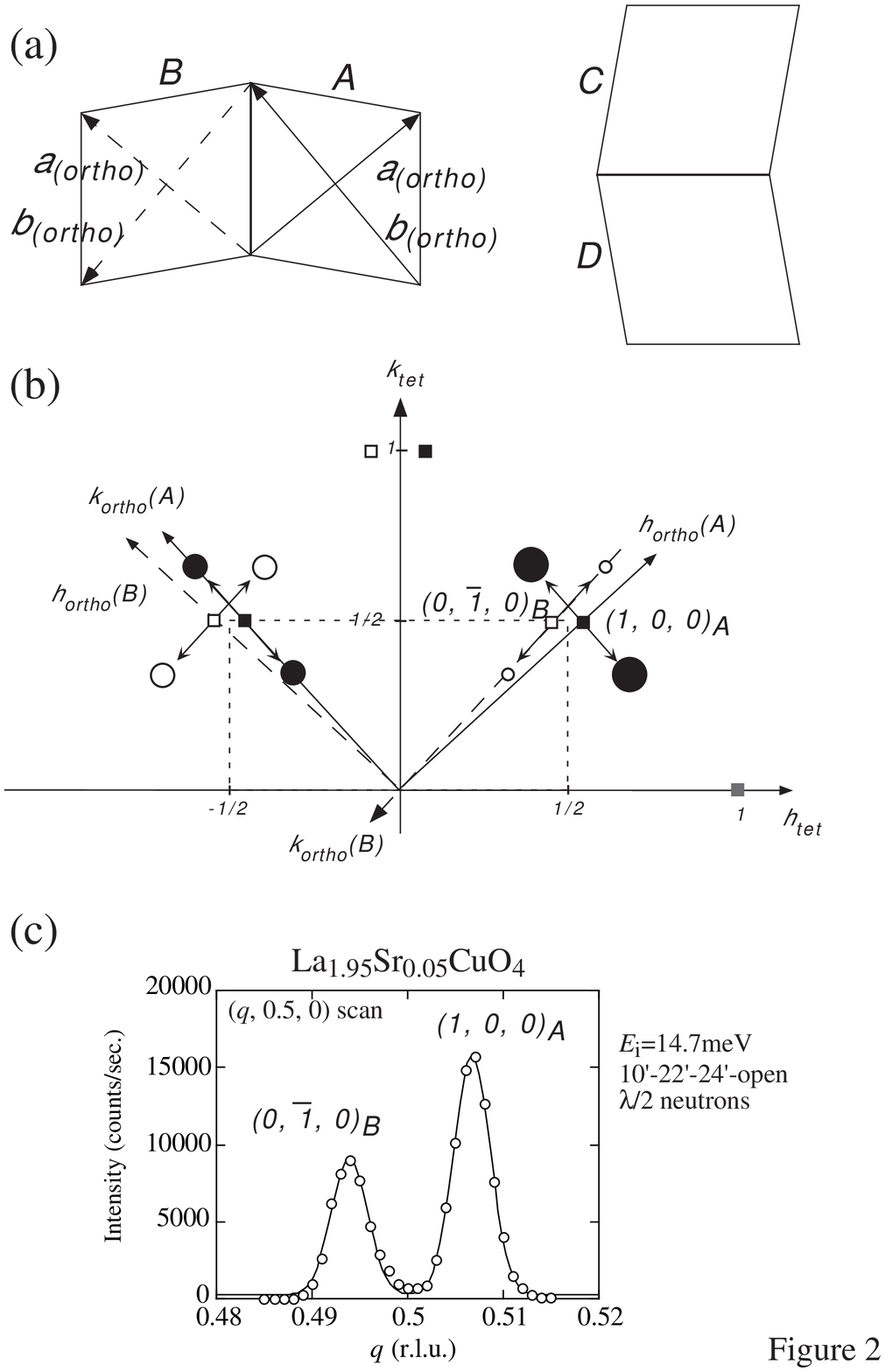}}
\caption{(a) Schematic drawings of the CuO$_{2}$ planes in the LTO structure.
The twinned structure normally includes these four structures.
(b) Peak positions of the IC magnetic peaks together with the fundamental Bragg peaks.
Circles and squares indicate the positions of the IC magnetic peaks and Bragg peaks, respectively.
Note that the Bragg peak positions are observed via $\frac{\lambda}{2}$ neutrons.
Closed and open symbols show peak positions in the A and B twins, respectively.
The size of circles corresponds to the relative intensity.
(c) Profile along the $(q,\ 0.5,\ 0)$ line in tetragonal notation.}
\label{fig_geometry}
\end{figure}

The LSCO system exhibits a structural transition from the high-temperature tetragonal 
(HTT) structure $(I4/mmm)$ to the low-temperature orthorhombic (LTO) phase $(Bmab)$. 
At the transition temperature, the CuO$_{6}$ octahedra coherently tilt 
along either the $[1 1 0]$ or the $[1 \bar{1} 0]$ direction in the $I4/mmm$ notation.
As a consequence, the squares of the CuO$_{2}$ lattice distort into diamonds,
as illustrated in Fig.~\ref{fig_geometry}(a), 
and the $\overrightarrow{a}_{ortho}$ and $\overrightarrow{b}_{ortho}$ 
axes of the LTO structure are the diagonals of the diamonds. 
Note that the relation between the lattice constants of the HTT and LTO 
structures, $a_{tet}$ and $a_{ortho}$, is approximately $a_{ortho}=\sqrt{2}a_{tet}$. 
The in-plane lattice constants of the $x=0.05$ crystal in the present study are $a_{ortho}= 5.34$~\AA\  
and $b_{ortho}=5.41$~\AA\ at 2~K, so that $(b/a-1)=0.013$. 

Since the $[1 1 0]$ and $[1 \bar{1} 0]$ directions are equivalent in the HTT phase, the LTO 
phase typically develops a structure including the 4 possible twins shown 
as A to D in Fig.~\ref{fig_geometry}(a). 
In the previous work for $x=0.05$ of Wakimoto {\it et al.},~\cite{waki_rapid} 
this twin structure was not taken into account explicitly.
Nevertheless, they were able to demonstrate that the 
elastic IC peak positions are rotated by $\sim 45^{\circ}$ from those of 
superconducting samples.~\cite{S.W.Cheong_91} 
The square symbols in Fig.~\ref{fig_geometry}(b) indicate the positions of the fundamental Bragg 
peaks for the $x=0.05$ crystal; these were observed using $\frac{\lambda}{2}$ neutrons. 
These data demonstrate that the sample has only two twins; the closed and open 
squares correspond to A and B twins respectively. Furthermore, the relative 
population of the A and B twins is 2:1.  This is determined from the relative intensities 
of the $(1,\ 0,\ 0)_{A}$ and $(0,\ -1,\ 0)_{B}$ reflections in LTO notation as shown in 
Fig.~\ref{fig_geometry}(c). 
It is fortunate that the sample has such an imbalanced twin structure since 
this allows us to characterize the static IC peaks in each twin separately.
Figure ~\ref{fig_geometry}(b) also summarizes the positions of the elastic IC peaks in the A and B 
twins which are indicated by closed and open circles, respectively. 
Once this orthorhombic twin structure has been established, the only remaining adjustable parameter 
is the ratio of the intensity of the IC peak around $(1,\ 0,\ 0)$ to that around $(0,\ 1,\ 0)$, 
which is determined by the spin structure.

As will be discussed in detail in the next section, 
the IC satellites appear 
at $(1,\ \pm q_{s},\ 0)$ and $(0,\ 1 \pm q_{s},\ 0)$ of the LTO structure in each twin; that is,
the spin modulation is one dimensional and its direction is
along the orthorhombic $b^{*}$-axis.
This one-dimensional spin modulation is similar to that of Cr.~\cite{Fawcett} 

For the IC peaks observed in the superconducting samples, 
the incommensurability parameter $\delta$ is defined 
as the separation of the IC peaks from the $(1/2,\ 1/2)$ position, expressed in tetragonal
reciprocal lattice units $(1~{\rm r.l.u.} \sim 1.65~{\rm \AA}^{-1})$.
Since for $x=0.05$ the spin modulation is parallel to the orthorhombic $b^{*}$-axis,
the parameter $q_{s}$ is most conveniently expressed in orthorhombic
reciprocal lattice units $(1~{\rm r.l.u.} \sim 1.17~{\rm \AA}^{-1})$.
However, throughout this paper, we prefer to use $\delta$ as the incommensurability parameter 
to facilitate direct comparison of the incommensurability for $x=0.04$ and $0.05$ 
with the values for $x \geq 0.06$.
$\delta$ is related to $q_{s}$ by the following formula in the limit of $b/a=1$:
\begin{equation}
q_{s}=\sqrt{2} \times \delta. \label{eq_1}
\end{equation}

Neutron scattering experiments were carried out on 
\linebreak
\noindent
single crystals of La$_{2-x}$Sr$_{x}$CuO$_{4}$ 
with $x=0.04$ and $0.05$
grown by the travelling-solvent floating-zone method.~\cite{lee_98}
The details of the crystal growth and other characterizations of the samples will be published  
elsewhere.~\cite{waki_stat}
We used primarily the triple-axis spectrometer, SPINS, installed in the cold neutron guide hall at the NIST reactor.
Some limited data were also taken with the spectrometer HER at JAERI in Japan. 
We mainly utilized incident neutron energies of 5~meV and 3.5~meV.
The horizontal collimation sequence was typically 32'-40'-S-40'-open.
Pyrolytic graphite (002) was used as a monochromator and an analyzer.
Contamination from higher-order neutrons was essentially eliminated by using a Be-filter for $E_{i}=5$~meV
and a BeO-filter for $E_{i}=3.5$~meV.

\begin{figure}
\centerline{\epsfxsize=3in\epsfbox{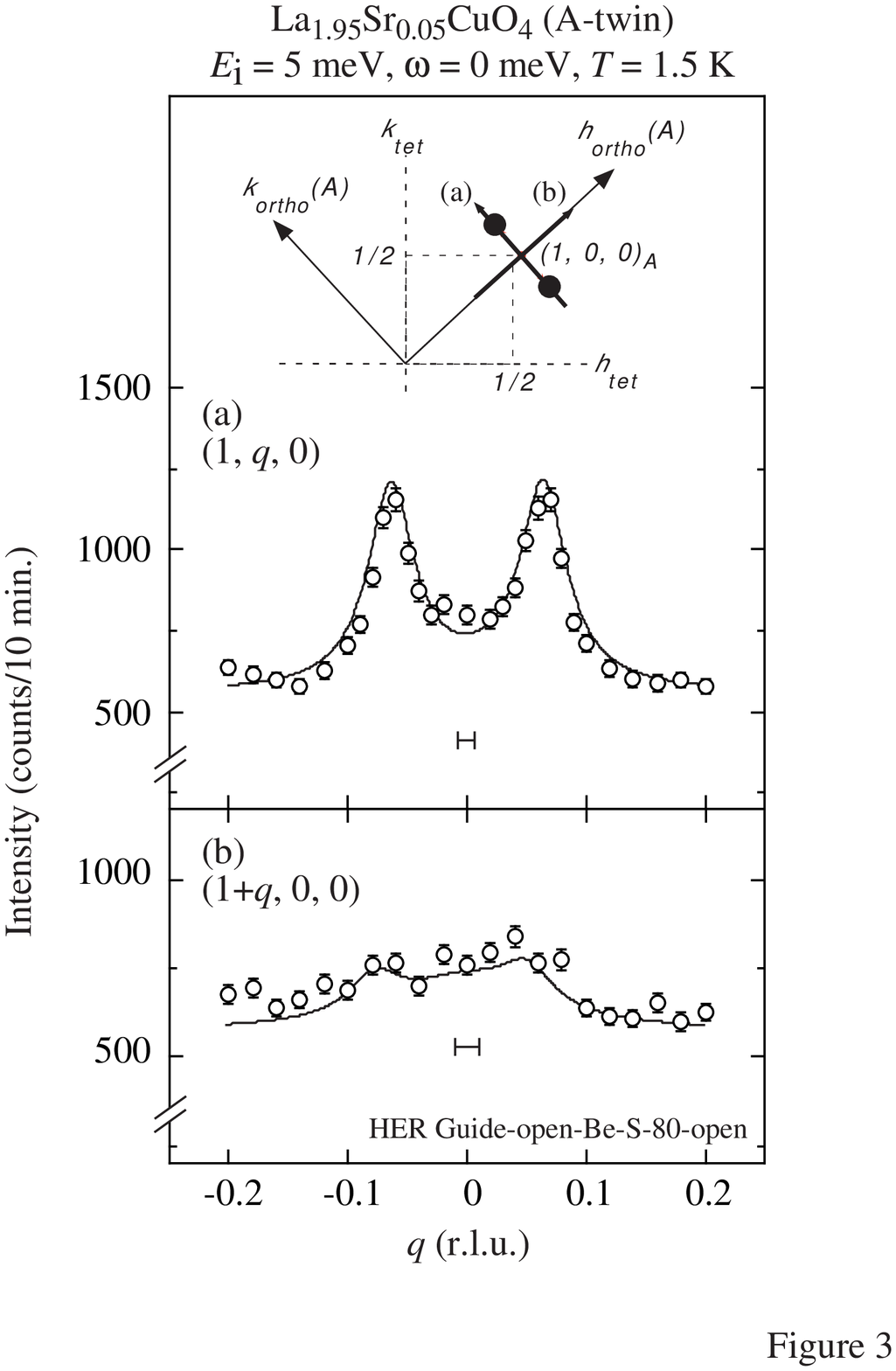}}
\caption{Peak profiles of the scans through $(1,\ 0,\ 0)_{A}$ along (a) the $b^{*}_{ortho}$ 
and (b) $a^{*}_{ortho}$ axes. Scan trajectories are illustrated in the inset. 
The solid lines are the results of the calculation as described in the text. 
The small horizontal bars indicate the instrumental resolution full width, $2\Gamma$.}
\label{fig_1stq}
\end{figure}

\section{Magnetic cross section}
\label{sec_result}

\begin{figure}
\centerline{\epsfxsize=3in\epsfbox{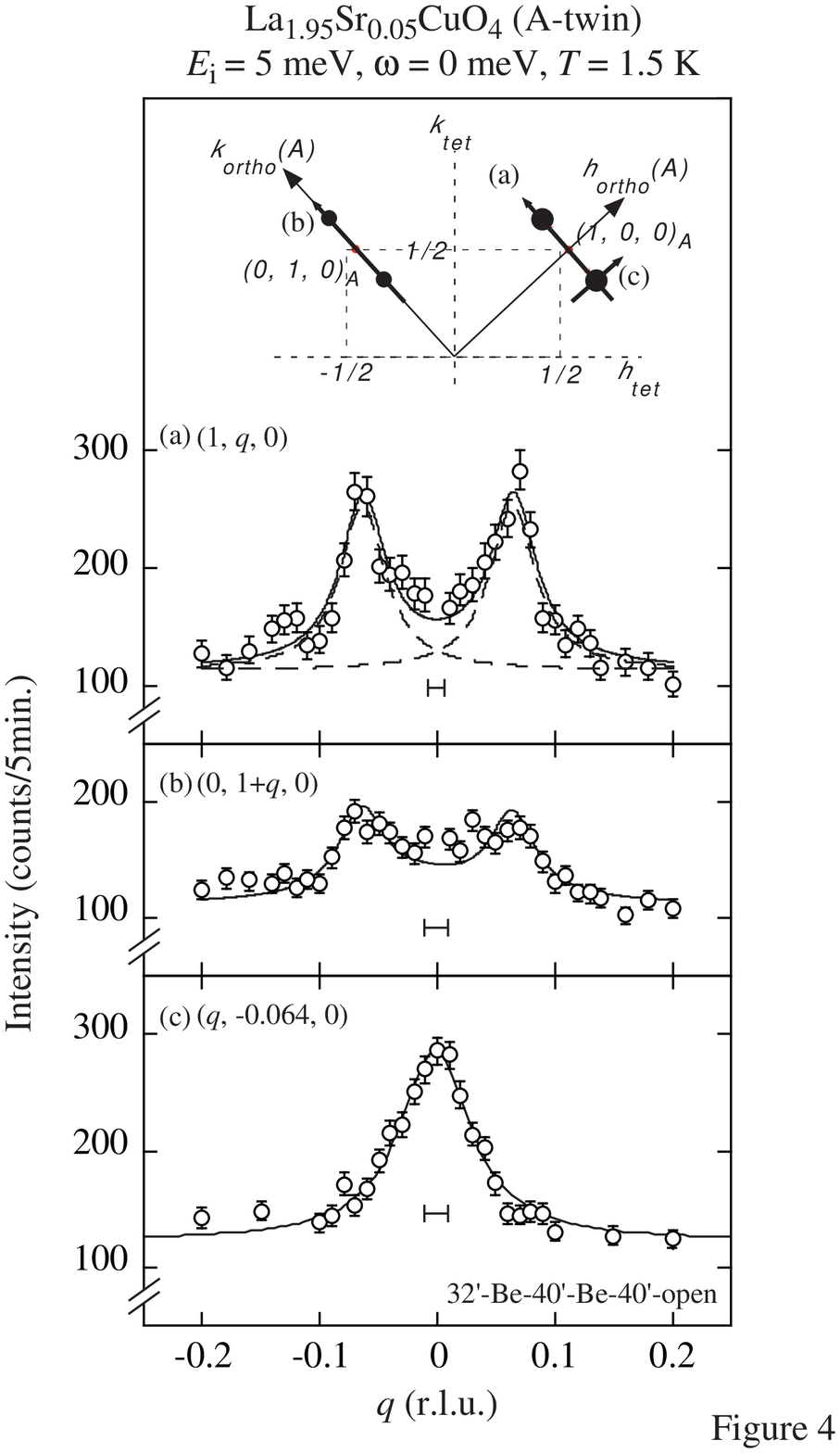}}
\caption{Magnetic cross sections of the A-twin. Scan trajectories are shown in the inset.
The solid lines indicate the results of the calculation as described in the text.
Dashed lines in Fig(a) show the individual IC peak components.
The small horizontal bars indicate the instrumental resolution full width, $2\Gamma$.}
\label{fig_Atwin}
\end{figure}

We measured the elastic magnetic peaks in each twin of the $x=0.05$ sample, separately.
At first, the spectrometer was set up so as to observe the $(H,\ K,\ 0)$ reflections of 
the A-twin; specifically, the solid lines in the insets of Fig.~\ref{fig_1stq} and \ref{fig_Atwin} 
were chosen as the $a^{*}_{ortho}$ and $b^{*}_{ortho}$ axes.
We show in Fig.~\ref{fig_1stq}, data taken on the spectrometer HER at JAERI.
The spectrometer configuration is given in the figure.
The data in Fig.~\ref{fig_1stq}(a) and (b) correspond to the scan trajectories (a) and (b)
illustrated at the top of Fig.~\ref{fig_1stq}.
Clearly the transverse scan (a) through $(1,\ 0,\ 0)_{A}$, that is, the (1,\ 0,\ 0) 
orthorhombic peak position for the A-twin shows well-defined peaks at $(1,\ \pm 0.064,\ 0)$.
On the other hand, there is barely any intensity observable above the background for 
the longitudinal scan through $(1,\ 0,\ 0)_{A}$.
In fact, the solid line in Fig.~\ref{fig_1stq}(b) is the calculated intensity due to the tails of the 
incommensurate scattering from the B-twin.
Thus, there is no observable intensity above the background along $a^{*}_{ortho}$ from the A-twin.
This means that the incommensurate spin modulation is one dimensional along $b^{*}_{ortho}$.
We now discuss data taken with SPINS at NIST.

Figure~\ref{fig_Atwin} shows the profiles of scans along the three trajectories shown 
at the top of the figure. 
As clearly demonstrated in Figs.~\ref{fig_Atwin}(a) and (b), pairs of satellite peaks exist 
around both $(1,\ 0,\ 0)_{A}$ and $(0,\ 1,\ 0)_{A}$ and the intensity ratio of the 
pairs around $(1,\ 0,\ 0)_{A}$ and $(0,\ 1,\ 0)_{A}$ is 2:1.
We believe that this difference in intensity arises from the spin orientation.
Possible spin structures which give consistency with this intensity ratio will be discussed 
in section \ref{spin_struc}.
A single peak at $q=0$ is observed in the profile along the trajectory of (c). 
These data together with those in Fig.~\ref{fig_1stq} provide direct evidence 
that the satellite direction is parallel to the $b^{*}_{ortho}$-axis.
This is also confirmed by the detailed numerical calculations discussed below.

\begin{figure}
\centerline{\epsfxsize=3in\epsfbox{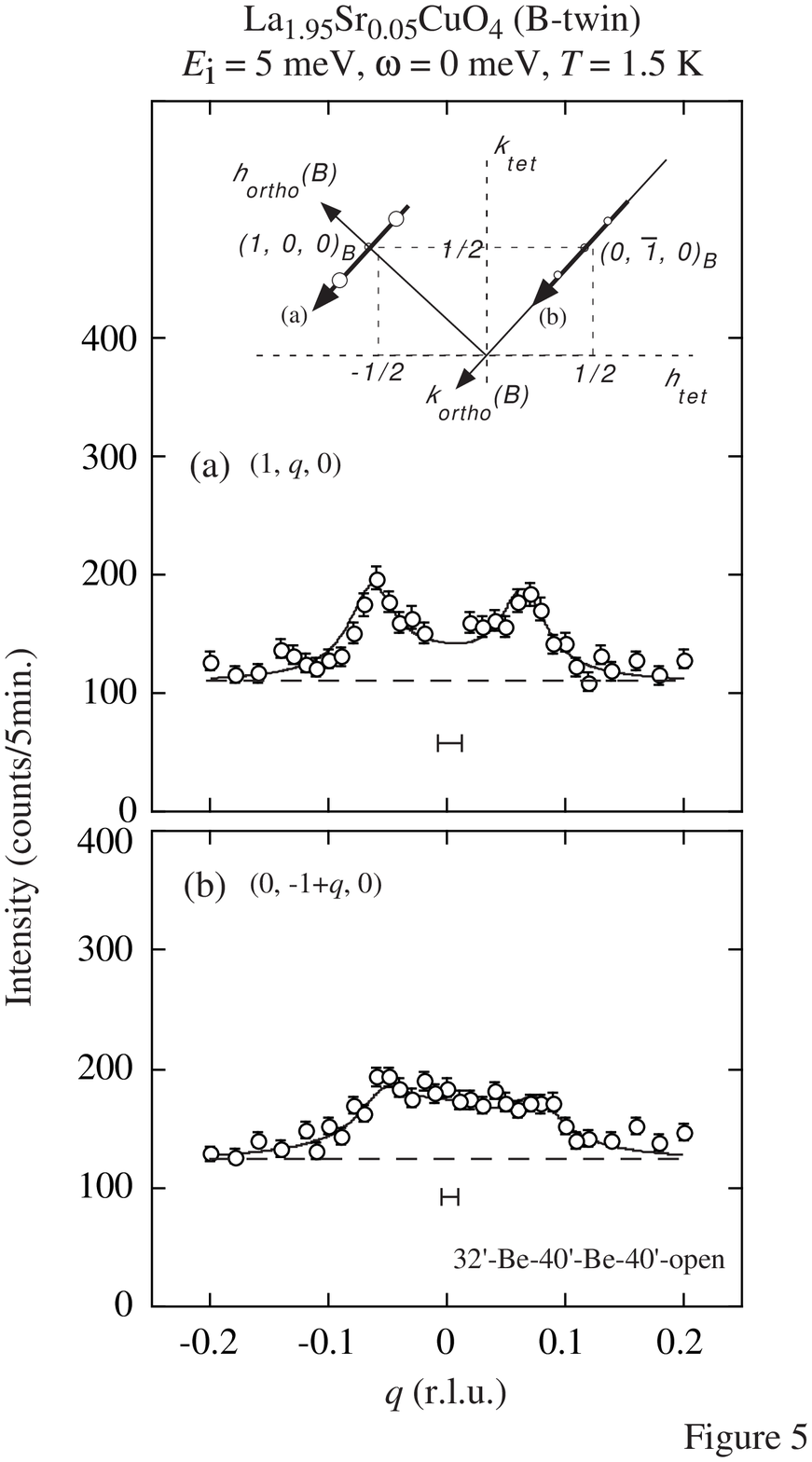}}
\caption{Magnetic cross sections of the B-twin. Scan trajectories are shown in the inset.
The solid lines indicate the results of the calculation as described in the text.
The dashed lines indicate the background level.
The small horizontal bars indicate the instrumental resolution full width, $2\Gamma$.}
\label{fig_Btwin}
\end{figure}

Next, the SPINS spectrometer was reset to observe the magnetic cross section of the B-twin; 
the solid lines in the inset of Fig.~\ref{fig_Btwin} were chosen as the $a^{*}_{ortho}$ and 
$b^{*}_{ortho}$ axes for twin-B.
Figure~\ref{fig_Btwin} shows the profiles along the two trajectories in the inset.
Although clear satellites 
\linebreak
\noindent
are observed around $(1,\ 0,\ 0)_{B}$ as shown in Fig.~\ref{fig_Btwin}(a), 
the IC peaks around $(0,\ -1,\ 0)_{B}$ are modified by the tail of the IC peaks coming from
$(1,\ 0,\ 0)_{A}$.

We have performed a numerical calculation and compared the results with the
measured scattering profiles to verify 
the IC spin structure.
From the experimental observation for the A-twin that the IC peak positions are $(1,\ \pm0.064,\ 0)_{A}$
and $(0,\ 1\pm0.064,\ 0)_{A}$, 
the IC peaks in the B-twin are accordingly expected to be located at 
$(1,\ \pm0.064,\ 0)_{B}$ and $(0,\ -1\pm0.064,\ 0)_{B}$.
The measured IC peak widths are typically 0.06~\AA$^{-1}$ full width at half maximum.
After deconvolution using a double Lorentzian function,
the intrinsic IC peak widths have been determined to be $\kappa_{b} \sim 0.03$~\AA$^{-1}$ and  
$\kappa_{a} \sim 0.04$~\AA$^{-1}$.
These correspond to real space distances of 33~\AA \ and 25~\AA \ respectively
which are quite small relative to those in, for example, 
La$_{2}$CuO$_{4.12}$ (Ref.15) where $\xi$ is greater than 600~\AA.
The intensity ratio of satellite pairs around $(1,\ 0,\ 0)_{A}$ and $(0,\ 1,\ 0)_{A}$ is 2:1 and
the relative population of the A and B-twins is 2:1. 
Therefore the intensity ratio of the four pairs around $(1,\ 0,\ 0)_{A}$, $(0,\ 1,\ 0)_{A}$, 
$(1,\ 0,\ 0)_{B}$ and $(0,\ -1,\ 0)_{B}$ should be 4:2:2:1.
We have calculated the two dimensional intensity distribution in the $(H,\ K,\ 0)$ zone
using the above parameters and including no commensurate component.
The results of the calculation are indicated as solid lines through the data in 
Figs.~\ref{fig_1stq}, \ref{fig_Atwin} and \ref{fig_Btwin}.
The profile in Fig.~\ref{fig_Btwin}(b) as well as the other profiles are all explained 
by our model. 
Furthermore, the asymmetry of the line shape in Fig.~\ref{fig_Btwin}(b) can be explained 
by the overlap of the small IC pair (symmetric for $q=0$) and the tail of 
the IC peaks coming from $(1,\ 0,\ 0)_{A}$ (centered at $q \sim -0.01$).
As stated above, these facts demonstrate that, for $x=0.05$, the spin system is modulated only along 
the orthorhombic $b$-axis in real space, and there is no commensurate component.

After the experiment on the $x=0.05$ sample was completed, we reexamined 
the $x=0.04$ sample, which was reported in Ref.9 to have a commensurate structure.
The motivation for this additional experiment was our realization that the particular 
macroscopic anisotropy of the magnetic cross section of the $x=0.05$ sample could give  
misleading results in the $(H,\ 0,\ L)$ zone.
This may be readily seen by  
looking at Fig.~\ref{fig_geometry}(b) and assuming that the $x=0.05$ crystal 
was mounted with the second quadrant 
$[110]_{tet}$ horizontal. 
This geometry would give a double peak in the $(H,\ 0,\ L)$ zone as reported in the previous 
paper for $x=0.05$,~\cite{waki_rapid}
since the intensities of the satellite peaks in the second quadrant are equivalent.
However, if the first quadrant $[110]_{tet}$ were mounted horizontal, 
then scans through $(1,\ 0,\ 0)_{ortho}$ would give a single peak. 
In fact, we have experimentally demonstrated this.
It turns out for the $x=0.04$ sample that 
the twin distribution is very similar to that of the $x=0.05$ sample.
By proper mounting of the sample we were able to observe 
well-defined IC peaks at $(1,\ \pm 0.052,\ L)$ and $(0,\ 1\pm0.052,\ L)$.
This is identical to the magnetic scattering geometry in the $x=0.05$ sample.
The incommensurabilities $\delta$ for the $x=0.04$ and $0.05$ samples are shown in Fig.1. 
The two data points for $x=0.04$ and $0.05$ are both consistent with the linear line $\delta \simeq x$.  

\begin{figure}
\centerline{\epsfxsize=3in\epsfbox{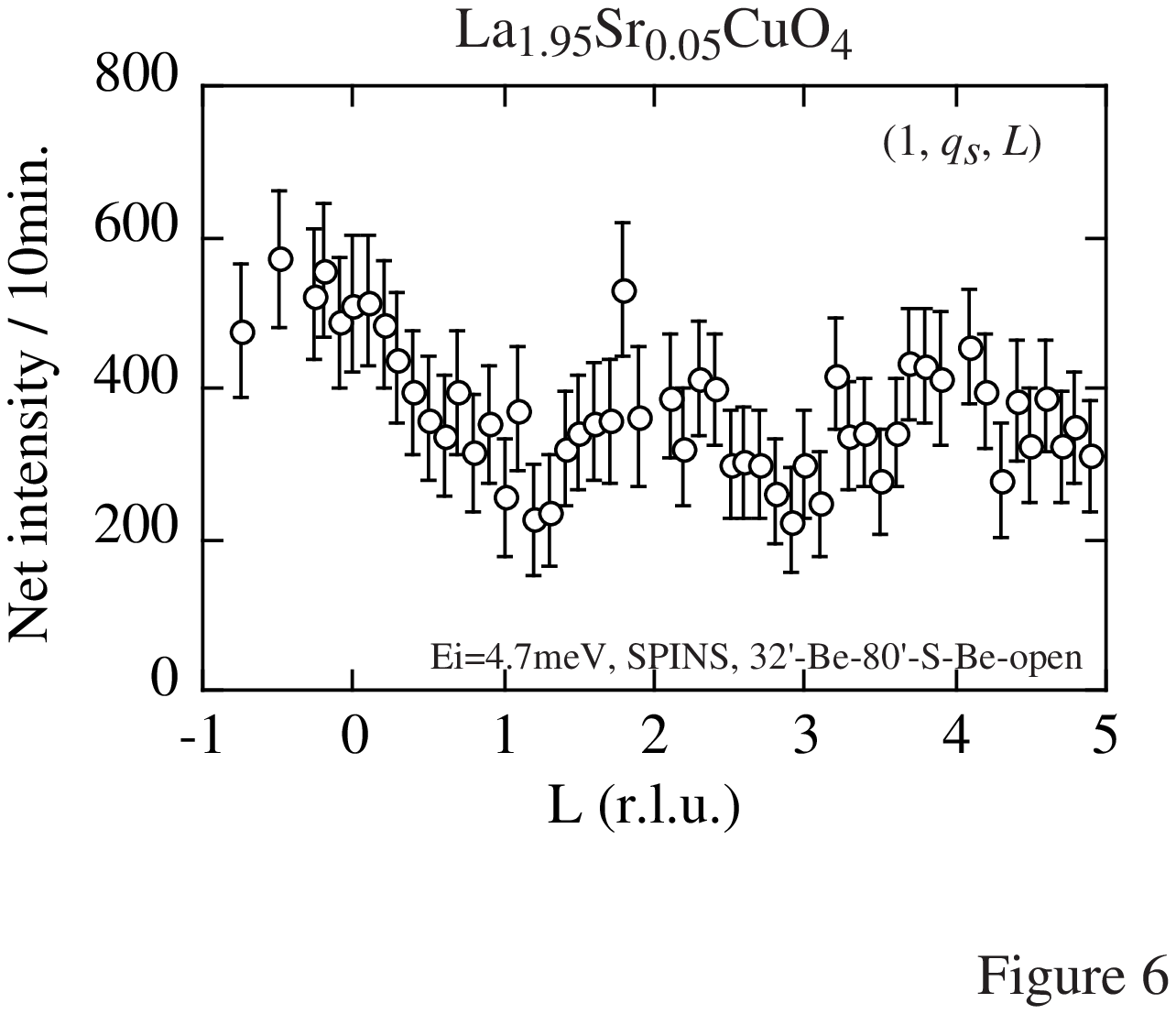}}
\caption{L-dependence of the magnetic scattering along $(1,\ q_{s},\ L)$ in 
La$_{1.95}$Sr$_{0.05}$CuO$_{4}$.
The data correspond to $I(T=1.5~{\rm K})-I(T=40~{\rm K})$.}
\label{fig_Ldep}
\end{figure}

Finally, as discussed previously in Ref.9, we have carried out a limited number of scans 
in the L-direction.
The results of these measurements are shown in Fig.~\ref{fig_Ldep}.
There is shown the difference between the measured intensity at 1.5~K and 40~K;
the resultant difference should be entirely magnetic
since over this temperature range any nuclear/structural neutron scattering should 
be independent of temperature.
We note that there is some contamination around $L \simeq -0.5$ which is not entirely removed 
by the subtraction process.
As may be seen in Fig.~\ref{fig_Ldep},
these measurements reveal only a weak L-dependence of the peak intensities, 
that is, the scattering
is essentially two dimensional.
This, in turn, means that the one dimensional spin modulations are only weakly correlated 
between successive CuO$_{2}$ planes.

\section{Discussion}

This paper reports the discovery of a new type of magnetic modulation for $x=0.04$ and $0.05$  
in the insulating range of LSCO at the boundary of the insulator-superconductor transition. 
This immediately raises the important question of how this new structure is related 
to the IC spin modulation collinear with the Cu-O-Cu-O bonds 
observed in the superconductors with $x=0.06$ or higher, 
as well as to the commensurate static and dynamic spin fluctuations observed in the
lower Sr concentration samples studied previously by
Thurston {\it et al.},~\cite{Bob_89} Keimer {\it et al.}~\cite{B.Keimer_92} 
and Matsuda {\it et al.}~\cite{Matsuda_99}
As mentioned in the previous section, we have reexamined the $x=0.04$ sample motivated by our own results
for $x=0.05$.
The measurements for $x=0.04$ demonstrate that the diagonal IC spin modulation phase is not 
limited to a narrow region right at the insulator-superconductor boundary.
An extensive study is now underway by Fujita {\it et al.} 
with special attention to concentrations at the boundary of the 
insulating to superconducting regions.
Preliminary results indicate that
the same type of diagonal IC magnetic structure is observed for $x=0.047$ and $0.053$.
These results will be published separately.
The next step of our investigation is, obviously, to probe the spin excitations 
in the diagonal spin density modulation state at low temperatures. 
For this purpose, it is essential to produce large untwinned crystals; this is now being attempted.. 
Here we comment briefly on some specific aspects of the spin correlations revealed by the current 
investigations.

\subsection{Polar coordinates}
\label{sec_polar}

Perhaps the most important result in the present work is that 
for temperatures below $\sim 15$~K in La$_{2-x}$Sr$_{x}$CuO$_{4}$ 
\linebreak
\begin{figure}
\centerline{\epsfxsize=3in\epsfbox{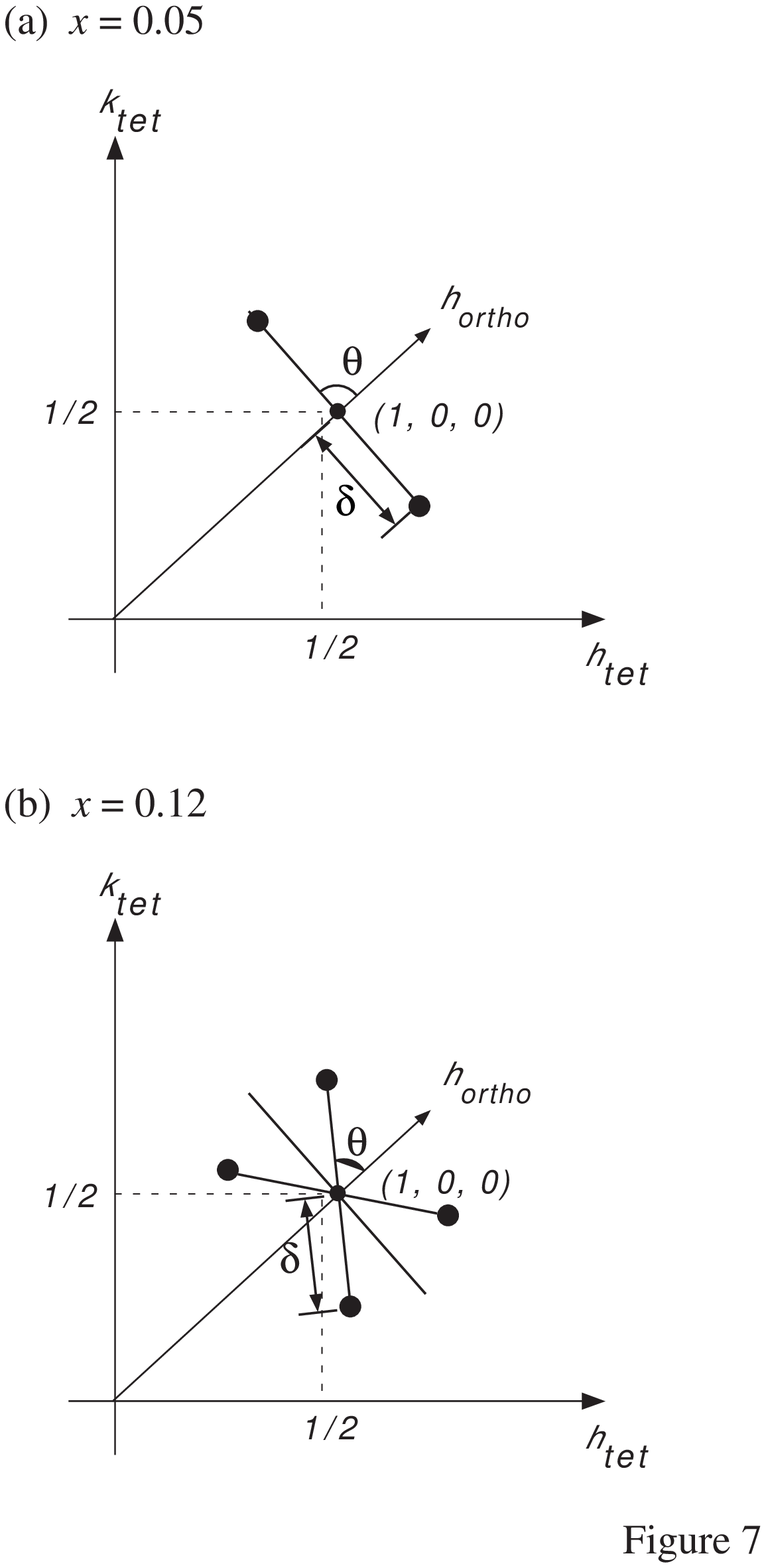}}
\caption{Schematic figure of the IC magnetic peak positions for (a) $x=0.05$ and (b) $x=0.12$.}
\label{fig_polar}
\end{figure}
\noindent
there exists only one pair
of satellites along the $b^{*}_{ortho}$-axis around each $(\pi,\ \pi)$ position for $x=0.04$ and $0.05$.
This leads naturally to a new parameterization of the IC peaks in the 
LSCO system over a wide range of Sr concentrations $x$.
We can express the positions of the IC peaks in reciprocal space in terms
of polar coordinates $(r=\delta,\ \theta)$ as follows: 
in the insulating region $x \stackrel{<}{\sim} 0.05$, two satellites are located at
$\theta = 90^{\circ}$ with respect to the $a^{*}_{ortho}$ axis 
with amplitude $\delta$ as shown in Fig.~\ref{fig_polar}(a). 
Recall that we defined $\delta$ as the distance between the IC peaks and the
orthorhombic $(1,\ 0,\ 0)$ position (or, in tetragonal units, the $(1/2,\ 1/2,\ 0)$ position), 
using, for convenience, the reciprocal 
lattice unit of the tetragonal structure. (1~r.l.u.~$\sim1.65$~\AA$^{-1}$)
At higher hole concentrations, the two satellites split into four peaks located at positions
corresponding to $\theta \sim \pm 45^{\circ}$ and $\pm 135^{\circ}$, 
again in the orthorhombic notation.
Of course, these four peaks observed in superconducting samples with $x \geq 0.06$
may correspond to two pairs of two peaks from putative separate one dimensional spin modulations.
Over the concentration range $0.04 \leq x \leq 0.12$ including both insulating 
and superconducting regions,
$\delta$ follows the simple linear relationship $\delta=x$ as clearly demonstrated 
in Fig.~\ref{fig_incomme}.
Recent studies for $x=0.12$ by Kimura {\it et al.}~\cite{Kimura_99} as well as 
precise measurements on
La$_{2}$CuO$_{4+\delta}$ 
by Lee {\it et al.}~\cite{Lee_99} have revealed that the satellite directions deviate
subtly from those identified by Cheong {\it et al.}~\cite{S.W.Cheong_91} as the 
tetragonal [100] and [010] axes as shown in 
Fig.~\ref{fig_polar}(b).
Specifically the axes of the spin density modulation in both La$_{1.88}$Sr$_{0.12}$CuO$_{4}$
and La$_{2}$CuO$_{4+\delta}$ are rotated by $\sim$ 3 degrees
from the tetragonal axes towards the $b^{*}_{ortho}$ axis.
However, we can describe such phenomena generally by using polar coordinates.
This description satisfies orthorhombic symmetry and, in general, there is no reason for $\theta$ 
to be exactly 45 degrees for $x=0.12$. 
In the same sense, $\theta$ does not have to be exactly 90 degrees for $x=0.05$,
that is, there is no symmetry reason for the stripes to be exactly diagonal. 
The experimental observations, however, prove that any deviation from 90$^{\circ}$
for $x=0.04$ and $0.05$ is very small.

\subsection{Spin structure}
\label{spin_struc}

\begin{figure}
\centerline{\epsfxsize=3in\epsfbox{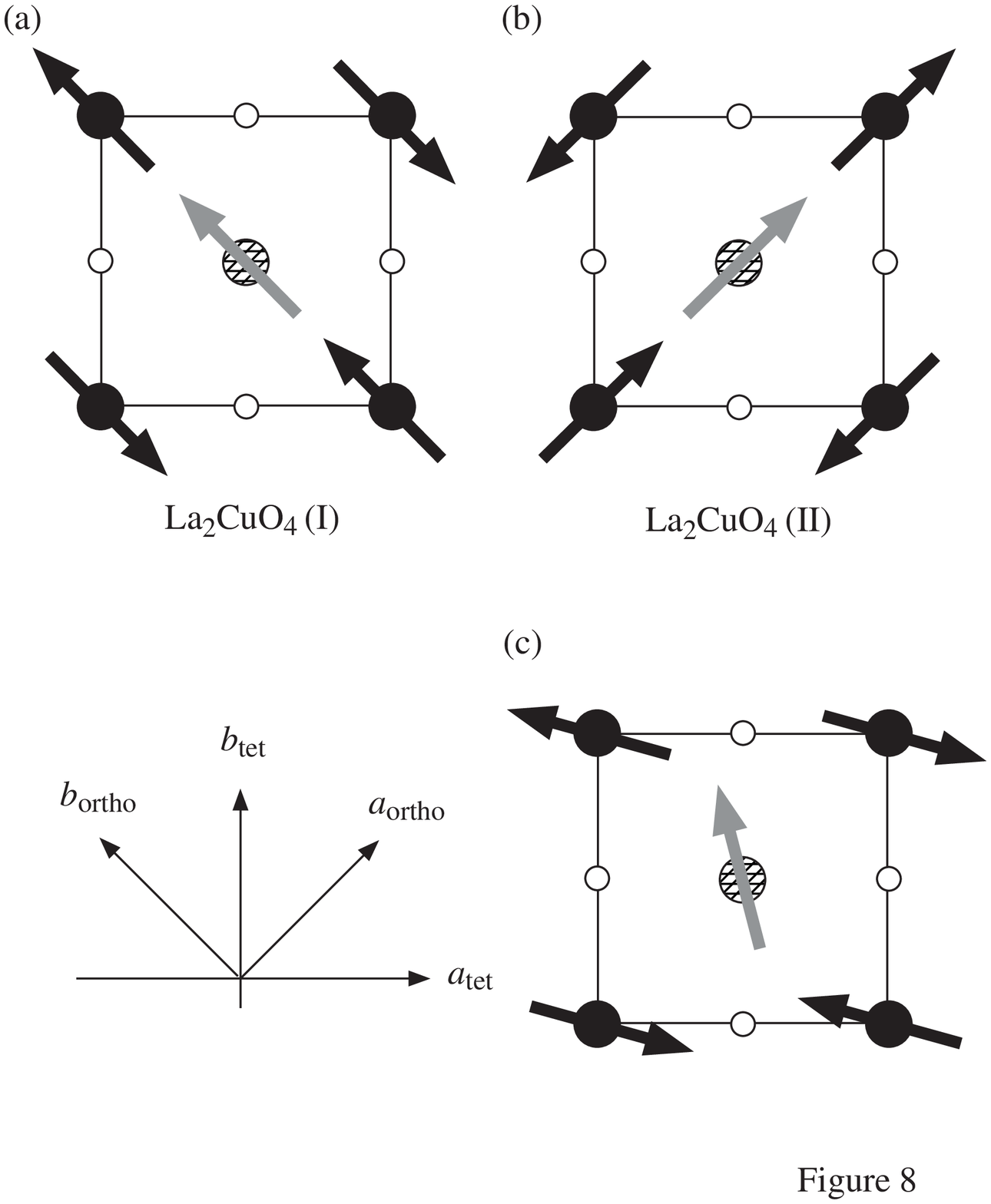}}
\caption{Spin structures of (a) La$_{2}$CuO$_{4}$(I), (b) La$_{2}$CuO$_{4}$(II) and 
(c) combined structure of those two in a single CuO$_{2}$ plane.
Closed circles indicate Cu atoms and open circles indicate oxygen atoms.
Shaded circles indicate the Cu atom on the next CuO$_{2}$ plane.
La$_{2}$CuO$_{4}$(I) and (II) structures give magnetic peaks at 
$(1,\ 0,\ 0)$ and $(0,\ 1,\ 0)$, respectively.}
\label{fig_spin}
\end{figure}

The $x=0.05$ sample shows IC peaks around both $(1,\ 0,\ 0)$ and $(0,\ 1,\ 0)$
with the intensity ratio of 2:1.
Figure \ref{fig_spin}(a) indicates the spin structure of undoped La$_{2}$CuO$_{4}$, 
in which the spin direction
is along the $b_{ortho}$-axis and the propagation vector is along the $a_{ortho}$-axis.
(Hereafter we call this magnetic structure LCO(I).)
This structure gives a magnetic Bragg peak at $(1,\ 0,\ {\it even})$, 
$(0,\ 1,\ {\it odd})$ and none at $(0,\ 1,\ 0)$.
A uniform rotation of the LCO(I) type does not create a $(0,\ 1,\ 0)$ peak
since the extinction rule is determined by the propagation vector not the spin direction.
One interpretation for the intensity ratio between the peak pairs around 
$(1,\ 0,\ 0)$ and $(0,\ 1,\ 0)$ is a broadening of the magnetic peaks 
along the $c^{*}$-direction due to a short magnetic correlation length along the $c$-axis
together with randomization of the spin direction;
that is, the intensity observed around $(0,\ 1,\ 0)$ is caused by tails of 
the peaks around $(0,\ 1,\ {\it odd})$.
According to our simulation, the LCO(I) structure with the correlation length 
of $\sim 8$~\AA \ and with a random spin orientation gives an intensity ratio
of 2:1 as observed experimentally.

Another interpretation for the intensity ratio is a mixture of two different spin structures 
with each component giving the magnetic peaks around $(1,\ 0,\ 0)$ and $(0,\ 1,\ 0)$ separately.
Figure \ref{fig_spin}(b) indicates one of the simplest structures which has $(0,\ 1,\ 0)$ 
as an allowed reflection.
In this structure (hereafter, referred to as LCO(II)), the
spin direction is parallel to the $a_{ortho}$-axis and the propagation vector is parallel to
the $b_{ortho}$-axis.
Based on the assumption that the magnetic peaks around (1, 0, 0) and (0, 1, 0) arise 
from the LCO(I) and LCO(II) components separately, 
possible models are
A) a mixture of 
\linebreak
\noindent
LCO(I) and LCO(II) structures with the volume ratio 2:1, 
and B) a non-collinear spin structure made by combining LCO(I) and LCO(II) 
as shown in Fig.~\ref{fig_spin}(c).
In the structure B), if one assumes a unique spin direction, then the spin 
must make an angle of $35^{\circ}$ with the $b_{ortho}$-axis.
However, from our simulation results, these models give a much narrower peak width 
along $c^{*}$-direction than that of the observed profile of Fig.~\ref{fig_Ldep}.
Thus the above "random spin orientation" model seems preferable.

According to magnetic susceptibility measurements,~\cite{waki_stat} the $x=0.05$ sample shows 
canonical spin-glass behavior with the spin-glass transition temperature, $T_{g} \sim 5$~K. 
At the same time our elastic or, strictly speaking, quasielastic neutron 
measurements reveal an onset temperature of 
$\sim 15$~K for the "elastic" incommensurate spin density 
wave peaks.~\cite{waki_rapid}
This difference in temperatures may simply reflect the differing energy resolutions of the 
various probes.
One possible explanation for the coexistence of a spin-glass state and a static one dimensional 
spin modulation 
is a cluster spin-glass model: the spin system forms 
two dimensional antiferromagnetically correlated 
clusters which have different spin directions.
The "random spin orientation" model discussed above also stands in the cluster spin-glass model
since the spin direction of each cluster should be random. 
If we apply the cluster spin-glass model with the spin structure model of A) or B), 
then each cluster should have 
a three dimensional spin structure whose spin direction is uniformly rotated from the structure of 
Figs.~\ref{fig_spin}(a), (b) or (c).
In all models, the width of the elastic peaks corresponds to the inverse of the cluster size.

\subsection{Structure and IC spin modulations}

Various explanations have been offered for the IC peaks in the LSCO system including, for example, 
stripe models,~\cite{Emery_94}
nesting of the Fermi surface,~\cite{Fukuyama_94} and impurity pinning 
effects due to the doped Sr.~\cite{Gooding}
The stripe model,
which predicts a one dimensional spin density wave modulation due to microscopic phase separation 
of the doped holes into lines of charge,
most naturally explains our new result that only a pair of satellites 
exists around $(\pi,\ \pi)$.
In this section we discuss the physical properties on the basis of the stripe model.
We emphasize, however, that we only have direct evidence for a one dimensional 
spin modulation.
We have not yet observed any evidence of charge ordering.

Since the spin modulation vector for the $x=0.04$ and $0.05$ samples is along only the $b_{ortho}$-axis, 
in the stripe model the associated lines of charge would be parallel to the $a_{ortho}$-axis;
that is, the lines of charge would be along the direction of uniform CuO$_{6}$ octahedron tilt.
This is similar to the case of La$_{1.48}$Nd$_{0.4}$Sr$_{0.12}$CuO$_{4}$~\cite{J.M.Tranquada_96}
although in that case the structural details are quite different.
The Nd, Sr co-doped material shows elastic magnetic satellites with the modulation vector parallel to 
the $a_{tet}$-axes in the low temperature tetragonal (LTT) phase; 
in the LTT phase, the CuO$_{6}$ tilt direction is uniform along the $a_{tet}$-axes.
Therefore, the lines of charge again appear to run along the direction of uniform CuO$_{6}$ octahedron tilt.
Tranquada {\it et al.}~\cite{J.M.Tranquada_96} have suggested that the charge stripes 
could be pinned by the 
corrugation of the CuO$_{2}$ planes resulting from the tilt of the octahedra,
thence leading to a suppression of superconductivity and an enhancement of the
elastic IC magnetic peaks for Sr doping near $x=0.12$. 
By analogy, in the case of La$_{2-x}$Sr$_{x}$CuO$_{4}$ with $x=0.04$ and $0.05$, 
the lines of charge along the $a_{ortho}$-axis could be pinned by 
the corrugation along the $b_{ortho}$-axis in the LTO phase 
resulting from the tilt of the octahedra.
At higher hole concentrations, $x \geq 0.06$, the charge stripes are rotated by $\sim$45$^{\circ}$ 
from those for $x=0.05$ in the LTO phase,
that is, they are in a direction of oscillating tilt;
this may help to explain why the system exhibits superconductivity for $x \geq 0.06$. 
It is still a completely open question why the charge line would change directions 
at $x \simeq 0.05$.
We should note, however, that some early theoretical calculations based on a simple Hubbard 
model~\cite{Schulz,Kato_90}
have predicted that charge stripes oriented along the $a_{ortho}$-axis should be stable at low 
hole concentrations.
Our new results suggest strongly that, in LSCO materials, 
the transport properties correlate directly with 
the relative orientations of the spin modulation vector and the CuO$_{6}$ tilt axis.
Since the CuO$_{6}$ tilt axis coincides with the direction of the Dzyaloshinski-Moriya
vector,~\cite{Thio}
this suggests that the magnetic anisotropy could play an important role in the magnetism 
and transport properties of the putative stripe phase.~\cite{Ortiz}

Finally, we discuss the hole-concentration dependence of the incommensurability $\delta$
in the presumed stripe model.
In the superconducting region, especially for $0.06 \leq x \leq 0.12$, $\delta$ follows 
the linear relation $\delta=x$ (Ref.8) as shown in Fig.~\ref{fig_incomme}.
For collinear stripes,
this relation corresponds to a hole concentration in a Cu-O-Cu stripe, $x_{r}$, of 0.5~holes/Cu.
Related behavior has been reported in the insulating 
La$_{2-x}$Sr$_{x}$Ni0$_{4}$ compounds.~\cite{T.traNi_96}
In the Ni system, the incommensurability $\epsilon$ obeys the linear relation $\epsilon=x$.
However, in that case $\epsilon$ has been defined as the distance between the IC peaks and 
the $(\pi,\ \pi)$
position in reciprocal lattice units of the {\it orthorhombic} structure
which, as mentioned in Sec.~\ref{sec_geometry}, differ by $\sim \sqrt{2}$ 
from the reciprocal lattice unit of the tetragonal structure.
It turns out that the relation, $\epsilon=x$, corresponds to $x_{r}=1$~hole/Ni.
In the present measurements for the $x=0.04$ and $0.05$ samples,
$\delta$ rather than $\epsilon$ follows well 
the $\delta=x$ line as clearly shown in Fig.~\ref{fig_incomme}.
This implies that the number of holes per unit
length on the stripe is constant through the diagonal to collinear stripe transition.
Since we defined $\delta$ using the reciprocal lattice unit of the {\it tetragonal} structure, 
the relation, $\delta=x$, corresponds to $x_{r}=0.7$~holes/Cu for diagonal stripes.
This means that the $x_{r}$ value may change suddenly at the insulator-superconductor boundary 
in the LSCO system.
We note that,
in the superconducting region, $\delta$ for $x=0.06$ is slightly lower than $x$
so that $x_{r}(0.06)$ is $\sim 0.6$~holes/Cu.
Therefore, it is possible that in the stripe model $x_{r}$ would evolve continuously 
from $\sim 0.7$~holes/Cu 
in the insulating region to $\sim 0.5$~holes/Cu in the superconducting region.
To elucidate this further, a much more 
detailed study is required around the
insulator-superconductor boundary with very small steps in $x$.

In brief, the salient
result of this study is that the insulator-superconductor transition in La$_{2-x}$Sr$_{x}$CuO$_{4}$
is coincident with a transition from a diagonal one dimensional {\it static} spin modulation
to a collinear spin modulation at low temperatures.
At the same time, as shown in the previous work,~\cite{B.Keimer_92}
the {\it instantaneous} magnetic correlations exhibit a commensurate-incommensurate
transition.
We believe that this rich magnetic behavior is central to 
the phenomenon of high temperature superconductivity.

\section{Acknowledgments}

We thank A. Balatsky, V.\ J.\ Emery, H.\ Fukuyama, K.\ Machida, M. Matsuda,
A. Millis, G. Ortiz, P.\ A.\ Lee, Q. Si and J.\ M.\ Tranquada for 
valuable discussions.  
R.J.B. thanks the Aspen Center for Physics for its hospitality during his part
in writing this paper.
The present work was supported by the US-Japan Cooperative Research Program on 
Neutron Scattering.  
The work at MIT was supported by the NSF under Grant No.\ DMR97-04532 and by the 
MRSEC Program of the National Science Foundation under Award No.\ DMR98-08941.  
The work at Tohoku and Kyoto has been supported by a Grant-in-Aid for Scientific Research of 
Monbusho and the Core Research for Evolutional Science and Techonology (CREST) 
Project sponsored by the Japan Science and Technology Corporation.  
The work at Brookhaven National Laboratory was carried out under Contract 
No.\ DE-AC02-98CH10886, Division of Material Science, U.\ S. Department of Energy. 
The work at SPINS in National Institute of Standards and Technology is based 
upon activities supported by the National Science 
Foundation under Agreement No. DMR-9423101.

\end{document}